\documentclass[8.5pt,twoside,twocolumn]{article}
\usepackage[super,sort&compress,comma]{natbib} 
\usepackage{mhchem}
\usepackage{times,mathptmx}
\usepackage{sectsty}
\usepackage{balance} 

\usepackage{graphicx} 
\usepackage{lastpage}
\usepackage[format=plain,justification=raggedright,singlelinecheck=false,font=small,labelfont=bf,labelsep=space]{caption} 
\usepackage{fancyhdr}
\begin{document}

\thispagestyle{plain}
\fancypagestyle{plain}{
\renewcommand{\headrulewidth}{1pt}}
\renewcommand{\thefootnote}{\fnsymbol{footnote}}
\renewcommand\footnoterule{\vspace*{1pt}%
\hrule width 3.4in height 0.4pt \vspace*{5pt}} 
\setcounter{secnumdepth}{5}

\makeatletter 
\def\subsubsection{\@startsection{subsubsection}{3}{10pt}{-1.25ex plus -1ex minus -.1ex}{0ex plus 0ex}{\normalsize\bf}} 
\def\paragraph{\@startsection{paragraph}{4}{10pt}{-1.25ex plus -1ex minus -.1ex}{0ex plus 0ex}{\normalsize\textit}} 
\renewcommand\@biblabel[1]{#1}            
\renewcommand\@makefntext[1]%
{\noindent\makebox[0pt][r]{\@thefnmark\,}#1}
\makeatother 
\renewcommand{\figurename}{\small{Fig.}~}
\sectionfont{\large}
\subsectionfont{\normalsize} 

\renewcommand{\headrulewidth}{1pt} 
\renewcommand{\footrulewidth}{1pt}
\setlength{\arrayrulewidth}{1pt}
\setlength{\columnsep}{6.5mm}
\setlength\bibsep{1pt}

\makeatletter
\DeclareRobustCommand\onlinecite{\@onlinecite}
\def\@onlinecite#1{\begingroup\let\@cite\NAT@citenum\citealp{#1}\endgroup}
\makeatother 

\twocolumn[
  \begin{@twocolumnfalse}
\noindent\LARGE{\textbf{CO$_{2}$/oxalate Cathodes as Safe and Efficient Alternatives \\ 
in High Energy Density Metal-Air Type Rechargeable Batteries}}
\vspace{0.6cm}

\noindent\large{\textbf{K\'aroly N\'emeth$^{\ast}$\textit{$^{a,b}$} and George Srajer\textit{$^{a}$} 
}}\vspace{0.5cm}


\noindent \textbf{\small{DOI: 10.1039/b000000x}}
\vspace{0.6cm}

\noindent \normalsize{
We present theoretical analysis on why and how rechargeable
metal-air type batteries can be made significantly safer and
more practical by utilizing CO$_{2}$/oxalate conversions instead of O$_{2}$/peroxide
or O$_{2}$/hydroxide ones, in the positive electrode.
Metal-air batteries, such as the Li-air one, may have very large energy densities, comparable to
that of gasoline,
theoretically allowing for long range all-electric vehicles.
There are, however, still significant challenges, especially
related to the safety of their underlying chemistries, the robustness of their
recharging and the need of supplying high purity O$_{2}$ from air to the battery.
We point out that the
CO$_{2}$/oxalate reversible electrochemical conversion is a viable alternative of
the O$_{2}$-based ones, allowing for similarly high energy density and almost identical voltage,
while being much
safer through the elimination of aggressive oxidant peroxides and the use of thermally stable,
non-oxidative and
environmentally benign oxalates instead.
}
\vspace{0.5cm}
 \end{@twocolumnfalse}
  ]

\section{Introduction}
High energy density batteries are expected to revolutionize transportation and allow for long-range
all-electric vehicles \cite{MArmand08,JChristensen12,GGirishkumar10,AKraytsberg11,FLi13}. 
The basis of exceptionally high gravimetric energy density metal-air batteries lies on one hand in the 
great free energy change of
metal-O$_{2}$ reactions, on the other hand in the fact that O$_{2}$ is available from air and does
not have to be carried on the vehicle. It has been pointed out in recent years that 
high energy density rechargeable Li-O$_{2}$ batteries can be built
when O$_{2}$ is converted to peroxide ions, O$_{2}^{2-}$ during the discharge of the battery \cite{TOgasawara06}. In
practice, such Li-O$_{2}$/peroxide batteries also produce Li-superoxide, LiO$_{2}$. Both of these discharge products are
aggressive oxidants that are difficult to control and may lead to unwanted side reactions. They may oxidize the
electrolyte and the porous carbon matrix often used as part of a gas-diffusion electrode in which the discharge product is
deposited. A high concentration of peroxides deposited in a carbon or general 
combustible matrix may lead to thermal runaway reactions, even to explosive combustions,
making such batteries a safety hazard. Other sources of hazards are posed e.g. by flammable electrolyte
components and very reactive dendrites on the surface of bulk lithium electrodes. 
Many of these hazards can be avoided 
by appropriate materials and techniques, discussed e.g. in Ref. \onlinecite{JChristensen12}. For example, all solid state Li-O$_{2}$ batteries
\cite{BKumar10} eliminate hazards related to dendrite formation and flammable electrolytes, however they continue to
deposit peroxide in a mix of carbon and ceramic material still leaving potential for explosive combustion of carbon.
Aqueous Li-O$_{2}$ batteries \cite{SJVisco10} produce lithium-hydroxide, LiOH, instead of Li$_{2}$O$_{2}$
in the cathode reaction 
O$_{2}$ + 2 H$_{2}$O + 4 e$^{-}$  $\to$ 4 OH$^{-}$. In these batteries the anode bulk lithium is immersed in an aprotic organic
electrolyte separated by a solid
Li-ion conducting ceramic membrane \cite{JFu97} from the aqueous solution of LiOH. 
The aqueous Li-O$_{2}$ battery has demonstrated
only limited or energy inefficient rechargeablity so far \cite{JChristensen12,TZhang13}. 
Aqueous electrolytes are also disadvantageous as they allow 
for the development of explosive hydrogen gas when operated outside their safe voltages windows
\cite{CWessells10}. 
For mechanical stability (e.g. avoiding punctuation) the Li-ion conducting membranes tend to be thick and
heavy, their ionic conductivity is low \cite{JChristensen12}.

Rechargeable metal-air batteries also have to use high purity O$_{2}$ to avoid the formation of large amounts of carbonates
due to the reaction of peroxides and hydroxides with CO$_{2}$ and that of nitrides in the case of non-aqueous metal-air
batteries. Such reactions would lead to the elimination of practical 
rechargeablity \cite{SRGowda13}, though for a primary, non-rechargeable battery, the addition of large amounts
of CO$_{2}$ ($\approx$ 20-80\%) in the gas feed, as an assist to O$_{2}$, leads to 2-3 fold increase of the realizable energy
density of such primary batteries \cite{KTakechi10,AKDas13}. The carbonates produced from a mix of 
O$_{2}$ and CO$_{2}$ feed require very large overpotentials during the charging, at least as long as no
catalyst will be found to reduce this overpotential to a practical value \cite{TZhang13}. The generation of pure O$_{2}$ from air,
free of H$_{2}$O, N$_{2}$ and CO$_{2}$, needed for metal-O$_{2}$ batteries without an on-board O$_{2}$-tank, represents a great
problem in itself. Unfortunately, large amounts of O$_{2}$ can only be stored in heavy tanks at high pressure, and 
the compression of O$_{2}$ requires such a large amount of energy that it renders batteries with on-board O$_{2}$-tanks
impractical \cite{JChristensen12}.

Additional problems arise from the storage of the discharge products. Solid discharge products deposited in the porous positive
gas diffusion electrodes clog the pores of the electrode and increase its internal electrical resistivity, 
thereby decreasing its energy storage capacity \cite{JChristensen12}. 
Ideally, discharge products would be removed from the space between the
electrodes and would be stored in a separate container, following the operating principle of flow batteries.
The optimal solvents for the dissolution of peroxides should allow for sufficiently fast ion transport for
fast discharge and charge, i.e. for high current densities. These solvents should also be safe, ideally non-flammable and
non-reactive with peroxides and whiskers on anodes. The best solvents developed for metal-air battery 
electrolytes appear to be blends of ionic
liquids (organic salts that are liquid at room temperature) and polar aprotic organic solvents \cite{BGKim13}.
Even with such solvents, explosive combustion of the electrolytes cannot be ruled out when 
the concentration of peroxides becomes large.

\begin{table}[tb!]
\small
\caption{
Standard enthalpies 
($\Delta_{f} H^{\ominus}$) and Gibbs free energies ($\Delta_{f} G^{\ominus}$) of formation of 
molecules and crystals relevant for energy storage reactions discussed in the present study.  
Gas and solid phases are referred to as (g) and (s), respectively.  The $\Delta_{f} G^{\ominus}$ 
value of Li$_{2}$C$_{2}$O$_{4}$ (s) has been calculated from the $\Delta_{f} G^{\ominus}$ values 
of Li(s) and CO$_{2}$(g) and the 3.0 V open circuit voltage of a Li-CO$_{2}$/oxalate cell based 
on the standard electrode potential of U$_{0}$(CO$_{2}$(g)/C$_{2}$O$_{4}^{2-}$) = -0.03 V from 
Ref. 
\onlinecite{RAngamuthu10} 
and the standard electrode potential of 
U$_{0}$(Li(s)/Li$^{+}$) = -3.04 V from Ref. 
\onlinecite{haynes2011crc} 
}
\label{FormationEnthalpies}
\begin{tabular*}{0.5\textwidth}{@{\extracolsep{\fill}}crrc}
\hline
 Molecule / Crystal        & \multicolumn{1}{c}{$\Delta_{f} H^{\ominus}$}  & \multicolumn{1}{c}{$\Delta_{f} G^{\ominus}$}  & Ref.     \\
                           & \multicolumn{1}{c}{(kJ/mol)}          & \multicolumn{1}{c}{(kJ/mol)}     &                     \\
\hline
C$_{8}$H$_{18}$(g)        &  -208.70       &      -347.95     & \onlinecite{WDGood72}     \\
O$_{2}$(g)                &     0.00       &       -61.17     & \onlinecite{MWJrChase98}  \\
H$_{2}$O(g)               &  -241.83       &      -298.13     & \onlinecite{MWJrChase98}  \\
CO$_{2}$(g)               &  -393.52       &      -457.26     & \onlinecite{MWJrChase98}  \\
Li(s)                     &     0.00       &        -8.67     & \onlinecite{MWJrChase98}  \\
Li$_{2}$O(s)              &  -598.73       &      -610.01     & \onlinecite{MWJrChase98}  \\
Li$_{2}$O$_{2}$(s)        &  -632.62       &      -649.44     & \onlinecite{MWJrChase98}  \\
Li$_{2}$CO$_{3}$(s)       & -1216.04       &     -1242.96     & \onlinecite{MWJrChase98}   \\
Li$_{2}$C$_{2}$O$_{4}$(s) & -1377.21       &     -1510.61     & \onlinecite{BBLetson65,RAngamuthu10}   \\
\hline
\end{tabular*}
\end{table}

\section{Results and Discussion}   
\subsection{Thermodynamics and Safety Considerations: Achieving High Energy Density and Improved Safety}
The conversion of O$_{2}$ to peroxides, such as Li$_{2}$O$_{2}$, instead of oxides or hydroxides,
is necessary to achieve practical rechargeability in metal-O$_{2}$ batteries 
\cite{TOgasawara06,JChristensen12}. 

\begin{table}[tb!]
\caption{
Standard reaction enthalpies ($\Delta_{r} H^{\ominus}$) and Gibbs free energies ($\Delta_{r} G^{\ominus}$)  
of energy storage reactions discussed in the present study, as calculated from 
$\Delta_{f} H^{\ominus}$ and $\Delta_{f} H^{\ominus}$ data of Table \ref{FormationEnthalpies}.
Data of the LiCoO$_{2}$-based Li-ion battery are from Refs. 
\onlinecite{KMizusima80}, \onlinecite{RYazami83} and \onlinecite{MMThackeray12}
}
\label{ReactionEnthalpies}
\begin{tabular}{ccrr}
\hline
System & Reaction                       &  $\Delta_{r} H^{\ominus}$ & $\Delta_{r} G^{\ominus}$       \\
       &                                &  \multicolumn{2}{c}{(kJ/mol)}            \\
\hline
n-octane              & C$_{8}$H$_{18}$(g) + 12.5 O$_{2}$(g) $\to$                       &           &   \\
                      &                     $\to$ 8 CO$_{2}$(g) + 9 H$_{2}$O(g)          & -5116     &  -5259     \\
Li-ion(LiCoO$_{2}$)   & Li$_{x}$C$_{6}$(s) + Li$_{1-x}$CoO$_{2}$(s) $\to$                &           &   \\
                      &                               $\to$ C$_{6}$(s) + LiCoO$_{2}$(s)  &    -      &  -213      \\
Li-O$_{2}$/oxide      & 2 Li(s) + 1/2 O$_{2}$(g)  $\to$                                  &           &   \\
                      &                                  $\to$  Li$_{2}$O(s)             &  -599     &  -562      \\
Li-O$_{2}$/peroxide   & 2 Li(s) + O$_{2}$(g)      $\to$                                  &           &   \\
                      &                                  $\to$  Li$_{2}$O$_{2}$(s)       &  -633     &  -571      \\
Li-[O$_{2}$+CO$_{2}$] & 2 Li(s) + CO$_{2}$(g) +                                          &           &   \\
 /carbonate           &  + 1/2 O$_{2}$(g) $\to$ Li$_{2}$CO$_{3}$(s)                      &  -823     &  -738      \\
Li-O$_{2}$/oxalate    & 2 Li(s) + 2 CO$_{2}$(g) $\to$                                    &           &   \\
                      &                           $\to$ Li$_{2}$C$_{2}$O$_{4}$(s)        &  -590     &  -579      \\
\hline
\end{tabular}
\end{table}
\begin{table*}[tb!]
\caption{
Theoretical gravimetric and volumetric energy ($\Delta_{r} G^{\ominus}$) densities and capacities of energy storage reactions, 
as well as open circuit voltages (OCV), densities of products and rechargeabilities.
$\Delta_{r} G^{\ominus}$ values have been taken from Table \ref{ReactionEnthalpies}. 
Note that the OCV for the Li-CO$_{2}$/oxalate cell is based
on the standard electrode potential of U$_{0}$(CO$_{2}$(g)/C$_{2}$O$_{4}^{2-}$) = -0.03 V from Ref. 
\onlinecite{RAngamuthu10} and on U$_{0}$(Li(s)/Li$^{+}$) = -3.04 V from Ref. 
\onlinecite{haynes2011crc}.
OCV-s of the other cells are based on Refs. 
\onlinecite{KMizusima80}, \onlinecite{RYazami83}, \onlinecite{TOgasawara06} and \onlinecite{MMThackeray12}.
The OCV of the Li-(O$_{2}$+CO$_{2}$)/carbonate cell is identical to that 
of the Li-O$_{2}$/peroxide one, as of Ref. \onlinecite{KTakechi10}, i.e. the addition
of CO$_{2}$ to a Li-O$_{2}$/peroxide cell produces extra heat instead of electrical energy,
while its rechargeability is debated \cite{KTakechi10,SRGowda13,TZhang13}.
O$_{2}$ or CO$_{2}$ may be supplied from air or from a gas tank carried on the vehicle leading to different energy
densities. 
Discharge capacities are referenced to bulk lithium and are identical for all Li-air type systems
(3830 mAh/kg and 2045 mAh/cm$^{3}$),
while charge capacities are referenced to solid discharge products.
Densities of solids are based on crystal structures at standard state 
}
\label{EnergyDensities}
\begin{tabular}{cccrrrrrrc}
\hline
System                &  OCV  & Density     & \multicolumn{4}{c}{Energy Density} & \multicolumn{2}{c}{Charge Capacity} & Recharge- \\
                      &   (V)     &of product   & \multicolumn{2}{c}{gravimetric} & \multicolumn{2}{c}{volumetric}& gravimetric &volumetric & ability      \\
                      &           &(kg/L) & \multicolumn{2}{c}{(Wh/kg)}     & \multicolumn{2}{c}{(Wh/L)} &(mAh/g) & (mAh/cm$^{3}$)  &              \\
                      &           &             & (air)   &  (tank) & (air)  & (tank)  &      &      &            \\
\hline
n-octane              &        -  &   -         &   12814 &   -     &  9008  &     -   &  -   &   -    &  N         \\
Li-ion (LiCoO$_{2}$)  &       3.6 &   5.05      &     -   &   568   &  -     &    2868 & 273  & 1379   &  Y         \\
Li-O$_{2}$/oxide      &       2.9 &   2.02      &   11151 &  5204   &  5955  &   10512 &  1787   & 3610   &  N         \\
Li-O$_{2}$/peroxide   &       3.0 &   2.25      &   11329 &  3448   &  6050  &    7758 &  1165   & 2621   &  Y         \\
Li-[O$_{2}$+CO$_{2}$]/&       3.0 &   2.10      &   11329 &  2143   &  6049  &    4500 &   724   & 1520   &  Y/N       \\
   /carbonate         &           &             &  + 3313 & + 627   & +1769  &  + 1316 &         &        &            \\
Li-CO$_{2}$/oxalate   &       3.0 &   2.14      &   11488 &  1577   &  6134  &    3375 &   525   & 1125   &  Y         \\
\hline
\end{tabular}
\end{table*}
The above mentioned issues with metal-O$_{2}$/peroxide batteries 
motivated us to seek alternative electrochemistries utilizing air-available species other than O$_{2}$ allowing for safer
and more practical high energy density batteries for long range all-electric vehicles.
Besides O$_{2}$, only CO$_{2}$ is an
air-available molecule with rich electrochemistry that has been studied extensively. 
The conversion of CO$_{2}$ to oxalate ions, according to the reaction 2 CO$_{2}$ + 2 e$^{-}$ $\to$
C$_{2}$O$_{4}^{2-}$, can be carried out with 100 \% selectivity and great
efficiency using existing catalysts \cite{RAngamuthu10} or properly chosen 
electrolytes/electrodes \cite{AKDas13,LSkarlos73}.
All other major studied reduction products of 
CO$_{2}$ would either involve hydrogen and would thus be less robust, 
or would lead to the evolution of poisonous carbon monoxide. 
While the deposition of
aggressive oxidant peroxides in combustible battery materials, such as carbon-based electrodes and electrolytes, may lead to
explosive combustions when peroxides are in high concentrations, there are no such problems with oxalates, as they are
non-oxidative, thermally stable and environmentally benign species. 
For example, Li$_{2}$C$_{2}$O$_{4}$ decomposes only at about 500 $^{o}$C \cite{DDollimore71}. Due to these features, oxalates are 
significantly safer discharge products than peroxides.
Oxalates are also known to be easily and quantitatively oxidizable to CO$_{2}$, in both aqueous and organic electrolytes
on various types of electrodes, including graphite \cite{BSlukic07,AKDas13,LSkarlos73,RAngamuthu10}.
Thus, reversible electrochemistries can be based on CO$_{2}$/oxalate
conversions. 

Remarkably, with the application of the copper-complex catalyst from Ref. \onlinecite{RAngamuthu10}, 
the CO$_{2}$ reduction standard electrode
potential becomes as high as U$_{0}$(CO$_{2}$(g)/C$_{2}$O$_{4}^{2-}$)=-0.03 V  
and the back-oxidation happens at U$_{0}$(C$_{2}$O$_{4}^{2-}$/CO$_{2}$(g))=+0.81 V \cite{RAngamuthu10},
leading to greatly reduced overpotentials and energy efficiency of the corresponding conversions.
Coupled with a Li-anode (U$_{0}$(Li(s)/Li$^{+}$)=-3.04 V, \cite{haynes2011crc}), 
this Li-CO$_{2}$/oxalate battery would have an open circuit voltage of $\approx$ 3.0 V, which is
practically identical with that of the Li-O$_{2}$/peroxide battery \cite{JChristensen12}. 
The identical voltages also imply close 
reaction Gibbs free energies ($\approx$ -575 kJ/mol), as both processes transfer two electrons per molecule of product. 

Tables \ref{FormationEnthalpies} and \ref{ReactionEnthalpies}
present the respective formation and reaction energies, while Table \ref{EnergyDensities} lists energy densities and
capacities of several Li-air type batteries in comparison with gasoline (n-octane) and a typical Li-ion battery. 
These data indicate that Li-CO$_{2}$/oxalate batteries may compete with Li-O$_{2}$/peroxide ones for applicability in
long-range all-electric vehicles. 
Indeed, the gravimetric energy density of the Li$_{2}$C$_{2}$O$_{4}$ formation is 11.5
kWh/kg, slightly greater than that of Li$_{2}$O$_{2}$ (11.3 kWh/kg), 
within 11\% to n-octane (12.8 kWh/kg), in reference to the weight
of Li, assuming air based O$_{2}$ or CO$_{2}$ intake. When O$_{2}$ or CO$_{2}$ is carried on the vehicle, or in reference to
the weight of the discharge products, the gravimetric energy density of Li$_{2}$C$_{2}$O$_{4}$ is 1.6 kWh/kg,
about 2.2 times smaller than that
of Li$_{2}$O$_{2}$, due to the larger weight of Li$_{2}$C$_{2}$O$_{4}$, still about 3 times larger than that of 
a LiCoO$_{2}$-based Li-ion battery.
The practical energy densities of the oxalate and peroxide based batteries would differ significantly less, 
as they would also involve the mass of other battery components, such
as electrolytes, membranes, current collectors, cases, etc. 
Furthermore, in case of the corresponding Na batteries (instead of
Li ones), the weight-factor would be reduced to 1.7 from 2.2. 
The gravimetric and volumetric capacities (Li-densities) of Li$_{2}$C$_{2}$O$_{4}$ are lower than those of 
Li$_{2}$O$_{2}$ by a factor of about 2.2. While the gravimetric capacity of Li$_{2}$C$_{2}$O$_{4}$ is about twice as much
as that of LiCoO$_{2}$, the volumetric one is only about 80 \% of it. 

The energy and capacity density values of Li$_{2}$C$_{2}$O$_{4}$ are quite practical to allow for long range electric vehicles.
A car that takes up 13 US (liquid) gallons (about 49 L) of n-octane,
stores about 88.6 kWh energy for useful work, assuming 20 \% engine efficiency. The equivalent of this energy
would be released by the production of 7.7 gallons of Li$_{2}$C$_{2}$O$_{4}$ at a rather poor
electric motor efficiency of 90 \%. 
The amount of oxalates needed may further be reduced with greater motor efficiency (96 \% as of Ref.
\onlinecite{TOikawa02}),  
electrically storing energy from braking or due to reduced idle time. 
Solvents for (partial) dissolution of the oxalates (or peroxides) may require additional space though.
The weight of 7.7 gallons of Li$_{2}$C$_{2}$O$_{4}$ is 62 kg, only about 27 kg more than that of 13 gallons of gasoline (35 kg).
The increase in the weight of the vehicle by the battery may be
reduced by lighter, carbon-composite based cars, and by electric motors lighter than internal combustion engines,
to maintain the same driving distance and approximately the same volume for
energy storage as known for present day gasoline driven cars.
 
The energy storage efficiency of the Li-CO$_{2}$/oxalate battery can be estimated from the voltage ratio of the discharge
and charge processes at constant current. 
Using the voltages of 3.0 and 3.8 V for discharge and charge, respectively, associated with the copper-complex catalyst of Ref.
\onlinecite{RAngamuthu10}, the energy storage efficiency of the Li-CO$_{2}$/oxalate
battery is estimated to be 79 \%. In practice, this value would be significantly lower but likely higher than that of
Li-O$_{2}$/peroxide
batteries, 65 \%, due to considerable overpotentials on charge in the latter ones \cite{AKraytsberg11}. 

In principle, a 1:1 molar mix of O$_{2}$ and CO$_{2}$ could lead to a further increased energy density in 
Li-[O$_{2}$+CO$_{2}$]/carbonate batteries, however, in practice either the O$_{2}$ or the CO$_{2}$ gets reduced, leaving
electrically utilizable energy densities (per mol of product) 
at the level of Li-O$_{2}$/peroxide or Li-CO$_{2}$/oxalate batteries.
When O$_{2}$ gets reduced, the presence of CO$_{2}$ leads to carbonate formation, producing a lot of additional heat,
formally through the reaction of Li$_{2}$O with CO$_{2}$ \cite{KTakechi10}. The
voltage of such a Li-[O$_{2}$+CO$_{2}$]/carbonate battery was found identical to that of the Li-O$_{2}$/peroxide one
\cite{KTakechi10}. The extra heat is disadvantageous as it 
will require additional cooling of the battery during discharge. It also indicates that a large amount of energy, 
one-third of that of the electrical energy stored, is needed to
free up CO$_{2}$ from carbonates on charge. These facts may render all metal-O$_{2}$/peroxide batteries inefficient
for rechargeable use, unless CO$_{2}$-free O$_{2}$-intake is applied \cite{SRGowda13}.
The current best rechargeable Li/ambient-air batteries produce mostly Li$_{2}$CO$_{3}$ during discharge
\cite{TZhang13} making the recharging process highly energy inefficient.

\begin{figure}[tb!]
\resizebox*{3.4in}{!}{\includegraphics{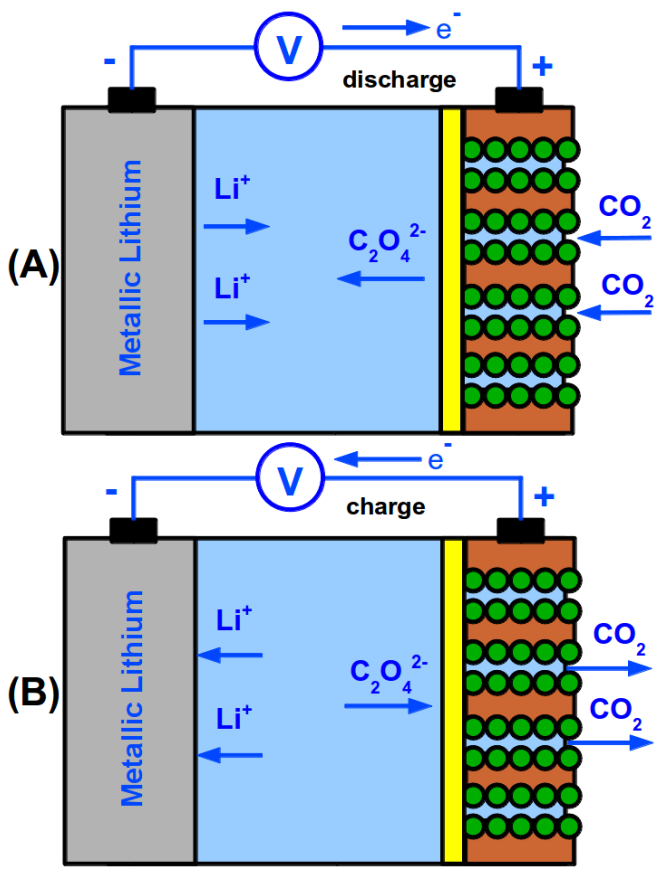}}
\caption{
Schematic view of one possible implementation of a Li-CO$_{2}$/oxalate battery. The positive electrode is made of
a porous electrically conducting material (brown color), 
which contains catalyst (green color) for reversible conversion of CO$_{2}$ to oxalate.
During discharge (panel A), oxalate ions (C$_{2}$O$_{4}^{2-}$) are produced from CO$_{2}$ diffusing in the cathode from
outside the cell. The oxalate ions migrate through the anion exchange
membrane (yellow color) and mix with Li$^{+}$ ions released from the Li-anode in the aprotic
electrolyte (blue color) that may be composed of a suitable ionic liquid. During the charging process (panel B)
Li is deposited in the negative electrode, while oxalate ions are converted back to CO$_{2}$ in the positive 
electrode and leave the cell.
}
\label{LiCO2oxalate}
\end{figure}

Interestingly, in some electrolytes based on ionic liquids, it is the CO$_{2}$ that gets reduced to
oxalate and O$_{2}$ does not formally participate in reactions \cite{AKDas13}.                                  
Selective reduction of CO$_{2}$ to oxalate
from ambient air through a catalyst \cite{RAngamuthu10} and simultaneous reduction of O$_{2}$ to peroxide 
is also possible, 
potentially allowing for a rechargeable Li-air battery by eliminating the presence of carbonates.

Pure O$_{2}$ or CO$_{2}$ could also be supplied from a tank stored on board of the vehicle. 
At a relatively safe and practical 120 bar pressure and 30 $^{o}$C temperature, the density of CO$_{2}$ is 0.802
kg/L, while that of O$_{2}$ is only 0.160 kg/L. With good thermal insulation, potentially even solid CO$_{2}$,
`dry ice', could be stored on board, with a high density of 1.6 kg/L. 
For the above mentioned 88.6 kWh useful energy, the space required for CO$_{2}$ storage would be between
8.8 and 18 gallons, while that of O$_{2}$ would be at 32 gallons. 
The compression of O$_{2}$ requires far more work though than that of CO$_{2}$. This is obvious also from the
economical availability of `dry ice', or the use of CO$_{2}$ as working fluid in air conditioning in both vehicles and
buildings \cite{US6588223}. 
Even though the work invested in the compression of these gases
may be returned to some extent when a higher pressure gas is applied on the respective electrodes 
\cite{JChristensen12}, CO$_{2}$ appears far more advantageous for on-board storage than O$_{2}$, for its far better
compressibility.
CO$_{2}$ can be collected efficiently from air as well, via CO$_{2}$-sponge materials, based on economically available 
ion-exchange resins \cite{KSLackner09,US7833328} allowing for CO$_{2}$ intake from air at rates of
0.25-0.83 g CO$_{2}$/m$^2$/s \cite{KSLackner09} (up to 3.0 kg CO$_{2}$/m$^2$/h)
despite the low, $\approx$ 0.04 mol \% concentration of CO$_{2}$ in air.

On charge, oxalate salts from metal-CO$_{2}$/oxalate batteries will be converted to CO$_{2}$ 
that will not react with other battery components, such as the
electrolyte or the porous carbon electrode, at the charging electrode potential.
Peroxide salts from metal-O$_{2}$/peroxide batteries will, however, be converted to highly reactive oxygen species  such as 
singlet oxygen \cite{JHassoun11} 
that may react with the electrolyte and the porous carbon or other combustible battery components.
While materials and techniques are being explored to avoid such unwanted side reactions during 
the charging of a metal-O$_{2}$/peroxide battery \cite{JChristensen12},
metal-CO$_{2}$/oxalate batteries completely eliminate this problem through the robustness
of quantitative oxidation of oxalates to CO$_{2}$ \cite{BSlukic07,RAngamuthu10}.

Figure \ref{LiCO2oxalate} shows one possible implementation of a simple Li-CO$_{2}$/oxalate battery \cite{US8389178}.
The porous positive gas-diffusion electrode contains catalysts, such as the copper-complex 
\cite{RAngamuthu10} mentioned above, for the reversible conversion of CO$_{2}$ to oxalate ions. During discharge,
CO$_{2}$ is converted to oxalate ions that migrate through an anion-exchange membrane \cite{US6716888} into the central
compartment of the cell where they mix with Li$^{+}$ ions released from a bulk Li-anode, while electrons move from the
anode to the cathode. In this implementation, the
aprotic electrolyte in the central compartment may be composed of a blend of an ionic liquid and a polar organic solvent, similarly
to that in Ref. \onlinecite{BGKim13}. The negative ions of the ionic liquid must exclusively be oxalate ions, 
so that no other negative ions get oxidized in the positive electrode on charge. The positive ions
of the ionic liquid may be quaternary ammonium ions, such as pyrrolidinium, imidazonium or 
tetrabutylammonium ones. 
Such ionic liquids have already been explored to some extent \cite{ARachocki08,US5371171}.
The polar organic solvent may be a mix of propylene carbonate and dimethoxyethane (monoglyme), similarly to electrolytes in 
Li-O$_{2}$/peroxide batteries, its main role is to dissolve the organic oxalate ionic liquid and
descrease its viscosity. This electrolyte may also be used in
the porous positive electrode to assist the transport of oxalate ions from/to the catalyst. 
Anion exchange membranes are used instead of cation ones to avoid the deposition of discharge
products in the positive electrode and thus minimize the amount of catalysts needed. The rate of oxalate ion transport
in the cathode, through the membrane and in the central compartment determines
the current density of the battery (rate of charge/discharge capacity) and should be subject of optimization. 
On charge, oxalate ions migrate to the positive electrode where they get oxidized back to CO$_{2}$, while
Li$^{+}$ ions get reduced and deposited on the negative electrode and electrons move from the positive electrode to the
negative one through the outer circuit. 
Many variants of the above described simple implementation of a 
Li-CO$_{2}$/oxalate battery are possible and some are discussed to a larger extent in Ref. \cite{US8389178}.

\subsection{Kinetics Considerations: Optimization of the Power Density}

In order to maximize the power density, i.e. the rate of charging/discharging the battery, the kinetic processes involved
in the electron transfers and ion transports have to be made quick. The application of CO$_{2}$ instead of 
O$_{2}$ offers several specific advantages in this respect as well.

\begin{figure}[tb!]
\resizebox*{3.4in}{!}{\includegraphics{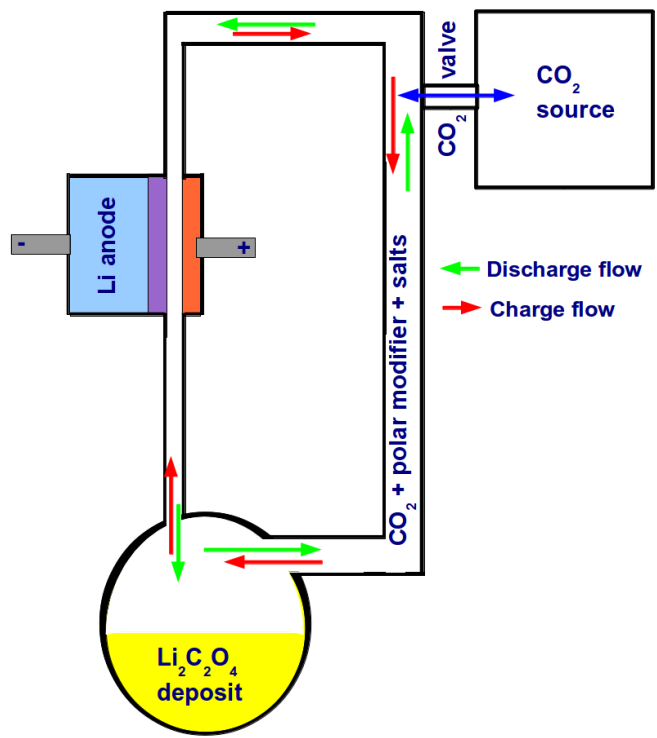}}
\caption{
Schematic view of a rechargeable Li-CO$_{2}$/oxalate flow battery. 
The purple stripe refers to a Li-ion selective membrane, the orange one to the catalytic
CO$_{2}$/oxalate electrode.
The arrows indicate the flow of the electrolyte during charge and
discharge. For maximum charge/discharge rates a pressurized solution or supercritical fluid 
electrolyte is applied consisting of high pressure (10-140 bar) mixture of CO$_{2}$, 
polar modifier (propylene carbonate,
glyme, etc) and supporting electrolyte (organic salt, ionic liquid). 
As the electrolyte passes between the protected Li anode
and the catalytic CO$_{2}$/oxalate cathode, it dissolves the discharge 
product Li$_{2}$C$_{2}$O$_{4}$ and deposits it in the product
container either through locally expanding volume and reduced pressure, or through a filter. 
The flow of the electrolytic fluid is recompressed after the deposition of 
Li$_{2}$C$_{2}$O$_{4}$ and circulated back toward the electrodes.
On charge, the flow of the electrolyte is reversed, 
Li$_{2}$C$_{2}$O$_{4}$ is dissolved in the electrolyte in the
product container and is converted back to Li and CO$_{2}$ on 
the electrodes. The increased pressure causes CO$_{2}$ to
migrate back to the CO$_{2}$ container through a CO$_{2}$ selective membrane and valve. 
A similar flow battery may be realized at normal pressure as well, 
utilizing deep eutectic solvents as electrolytes
based on organic and inorganic oxalates, as discussed in the text.
The CO$_{2}$ source unit may be a CO$_{2}$ tank (for high pressure applications) or an
atmospheric CO$_{2}$ absorbing unit (for normal pressure applications).
}
\label{SupecritCO2Battery}
\end{figure}

In case CO$_{2}$ is taken from a tank and is supplied as a moderately high pressure (10-140 bar) 
gas or supercritical fluid into the battery, 
it can serve both as a solvent and a source of electroactive species in the same time. 
Fundamental aspects of electrochemistry in supercritical CO$_{2}$
have been investigated by Abbott et al. \cite{APAbbott96,APAbbott00}. The use of high pressure or 
supercritical CO$_{2}$ appears attractive
for achieving very fast ion transport and thus high charge/discharge rates, as diffusion may be very fast in such
high-pressure or supercritical
medium when the concentration of voids is large enough \cite{APAbbott04}. 
As liquid or fluid CO$_{2}$ has low solubility for salts, a polar modifier needs to be added, such as
1,1,1,2-tetrafluoroethane (HFC 134a) \cite{APAbbott00}, propylene carbonate \cite{PMantor82,HKawanami03} or glymes
\cite{DKodama11}. 
Both propylene carbonate and glymes are excellent solvents (or solutes) 
for (or in) pressurized CO$_{2}$ and mix well with supercritical CO$_{2}$ fluid. 
Propylene carbonate at room temperature and 10 to 55 bar pressure can disolve 
10 to 60 mol\% CO$_{2}$, respectively \cite{PMantor82}, while diglyme at 40 $^{o}$C and 10 to 71 bar pressure dissolves
20 to 85 mol\% CO$_{2}$, respectively \cite{DKodama11}. 
Note that the critical point of pure CO$_{2}$ is at
T$_{c}$=31 $^{o}$C and p$_{c}$=74 bar.
The compression of the pressurized/supercritical electrolyte comsumes a fraction of the
energy stored in the battery, however, this fraction can be kept small ($<$3\%) for isothermal
compression and even smaller when electrochemical compression is used,
as discussed by Christensen et al. in Ref. \onlinecite{JChristensen12} for Li-O$_{2}$
batteries.

In addition to the polar modifier,
supporting electrolyte is also needed to increase the ionic conductivity of the electrolyte, at least until it is
saturated with discharge products. The supporting electrolyte is typically composed
of organic salts (ionic liquids), such as tetrabutylammonium tetrafluoroborate  \cite{APAbbott00},
or imidazolium salts \cite{HKawanami03} and may be composed of the oxalate ion containing organic salts mentioned above,
as well. Additionally, many organic salts show
anomalous melting point depression in pressurized CO$_{2}$ \cite{AMScurto06,AMScurto08} 
which in principle allows such pressurized room
temperature ionic liquids to be used as electrolytes in themselves,
i.e. without the glymes or organic carbonates or other polar modifiers
while also maintaining increased concentration of voids for faster ion transport.

A flow battery based on pressurized/supercritical CO$_{2}$ is depicted in Fig. \ref{SupecritCO2Battery}. 
It is composed of a negative electrode with protected metal (Li, Na, Mg, Al, etc) source, a positive electrode 
current collector with a catalyst for CO$_{2}$ reduction and oxalate oxidation covering its surface, and a flow of
CO$_{2}$-rich solution or supercritical fluid between the electrodes. 
As the flow passes between the electrodes it becomes enriched in
oxalate salts during discharge. The oxalate salts will be deposited in a container where the pressure may decrease due to
increased volume and the CO$_{2}$ and the polar modifier or solvents will be recompressed and circulated back to between the
electrodes. In case a non-compressible solvent is used, it may be pumped through a filter leaving the
excess discharge products in the product container. 
On charge, the flow of the supercritical fluid or that of the pressurized solution will be reversed, 
oxalate ions will be oxidized back to CO$_{2}$,
metal ions reduced and the regenerated CO$_{2}$ will be stored back in the CO$_{2}$ tank.
The charge/discharge rates of such a battery are expected to primarily depend on 
the catalyst density on the surface of the positive electrode,
as the concentration of the CO$_{2}$ is very high and transport of electroactive species 
to and from the catalyst is expected to be very fast
due to the directed flow of the voids-rich fluid or liquid. 
The rates also depend on the ionic conductivity of the protective skin 
(or cation selective membrane) on the negative electrode. Highest discharge rates are expected for supercritical fluid
electrolytes for the greatest concentration of voids supporting ion transport.

In case CO$_{2}$ is not supplied from a tank but taken up from the air through a CO$_{2}$
absorbing material mentioned above, 
a similar flow battery scheme can be utilized, with a
circulated electrolyte at atmospheric pressure. 
In order to maximize the speed of transport of oxalate ions, the use of deep
eutectic solutions is suggested as primary component of the electrolyte, as it opens the 
way to hopping based fast oxalate ion 
transport as opposed to slow simple diffusion. It is expected that the mix of oxalate ion containing 
solid organic salts (such as those mentioned above with pyrrolidinium, 
imidazonium or tetrabutylammonium cations) and the
discharge product metal-oxalates forms room temperature liquids based on principles of 
deep eutectic mixtures. Examples of such deep
eutectic liquids with oxalate salts exist, and provide alternatives to ionic liquids as solvents. 
For example, the mixture of choline
chloride and oxalic acid forms a room temperature liquid electrolyte, even though the individual components are solids 
themselves. Electrochemistry in such deep eutectic solvents with carboxylic 
acids has been studied by Abbott et al. and by 
LeSuer et al. \cite{APAbbott04,CANkuku07}. The ionic conductivity of such deep eutectic liquids was found to be similar
to that of ionic liquids \cite{APAbbott04}. In the proposed mix of organic oxalates and metallic oxalates it is expected
that the oxalate ions would chelate the metal cations and oxalate transport would happen through
hopping of oxalate ions between neighboring chelating sites, instead of diffusion. 
Since the concentration of the chelating sites is very high in the proposed deep eutectic liquids, a fast oxalate ion
transport is expected. Note, that only the transport of oxalate ions is expected to be fast in such liquids, other anions
would be transported much slower.
Also note that the previously studied
choline chloride and oxalic acid mixture does not allow for hopping based oxalate transport as it does not contain metal cations
as chelating sites, thus the proposed hopping based oxalate ion transport is yet to be investigated experimentally.
The deep eutectic solvent may be diluted by the addition of polar aprotic solvents, 
such as propylene carbonate and
dimethoxyethane, to decrease its viscosity, especially for the fully charged state where the   
metal oxalate component may be missing as is yet to be produced during the discharge. 
When the deep eutectic liquid
becomes saturated with the dissolved metal oxalates, 
the precipitating excess oxalates would be deposited in the discharge
product container. On charge, the oxalates in the product container would be 
dissolved in the deep eutectic electrolyte
and brought to the electrodes where metal and CO$_{2}$ would be generated as 
the battery is being charged.
Also note that the cell in Fig. \ref{LiCO2oxalate} can also be  
used in a flow battery design, in this case the CO$_{2}$ is supplied through a gas diffusion
electrode separated by an oxalate ion conducting membrane from the electrolyte. 
The above described flow battery designs also have the advantage of minimizing the amount 
of the electrolyte and other additional components realtive to the reactants and the products.
Many variants of the CO$_{2}$/oxalate battery are possible utilizing the principles
described here. 

A primary battery with Na anode and CO$_{2}$/oxalate cathode has 
recently been realized by Das et al. \cite{AKDas13} as an unexpected by-product
of Na-(O$_{2}$+CO$_{2}$) battery experiments,
observing that oxygen admixture to CO$_{2}$ can 
act as a catalyst for the production of oxalate salts during discharge in 
appropriately chosen ionic liquid electrolytes while
no oxide or carbonate discharge products are formed. 
The voltage of this battery ($\approx$2.25 V, OCV) 
is about 18\% smaller than that of an analogous 
Na-(O$_{2}$+CO$_{2}$)/(peroxide+carbonate+oxalate) battery presented in the same work 
($\approx$2.75 V, OCV, see Fig. 1 of Ref. \onlinecite{AKDas13}) or that of 
a Na-O$_{2}$/(peroxide) battery (2.6 V, theoretical OCV).
This experimental result indicates the feasibility of metal-air type batteries with CO$_{2}$/oxalate
cathodes, and also indicates that oxalate formation may be similarly energetic 
as peroxide formation when proper catalysts and electrolytes are used.
This observation is in agreement with what we
described above about the thermodynamics of CO$_{2}$/oxalate conversion using experimental data 
with the copper complex catalyst 
by Angamuthu et al. \cite{RAngamuthu10} and the measured enthalpies and Gibbs free energies of 
formation of Li$_{2}$C$_{2}$O$_{4}$.

\section{Conclusions}
We have proposed to use CO$_{2}$/oxalate electrodes instead of O$_{2}$/peroxide or O$_{2}$/hydroxide ones
in high energy density rechargable metal-air type batteries. The advantages of
CO$_{2}$/oxalate electrodes may be realized in terms of 
significantly improved safety and environmental friendliness, 
high energy and power density, robust and more efficient rechargeability 
and efficient on-board storage or air-based intake of CO$_{2}$. 

\section{Acknowledgements} 
K. N. gratefully acknowledges helpful discussions with Drs. M. Balasubramanian, G. Crabtree, N. Markovic, L. Trahey, 
and M. van Veenendaal at Argonne National Laboratory and A. K. Unni, C. U. Segre and L. Shaw at IIT. This research was supported
by the U.S. DOE Office of Science, under contract No. DE-AC02-06CH11357.

\footnotetext{\textit{$^{a}$~Address, Advanced Photon Source, Argonne National Laboratory, 
Argonne, Illinois 60439, USA, nemeth@ANL.Gov}}
\footnotetext{\textit{$^{b}$~Address, Physics Department, Illinois Institute of Technology, 
Chicago, Illinois 60616, USA }}

\footnotesize{
\bibliographystyle{rsc} 

\providecommand*{\mcitethebibliography}{\thebibliography}
\csname @ifundefined\endcsname{endmcitethebibliography}
{\let\endmcitethebibliography\endthebibliography}{}

}

\end{document}